\documentstyle[11pt,paspconf]{article}
\font\sc=cmcsc10
\begin{document}

\title{Gravitational Lensing: Recent Progress and Future Goals -- Conference Summary}
\author{Paul L. Schechter}
\affil{Massachusetts Institute of Technology, Cambridge MA 02139}

\begin{abstract}
There are three neighborhoods of interest in gravitational lensing:
that of the source, that of the lens and that of the observer.  Since
the last major meeting on lensing, the 1995 IAU Symposium, No.\ 173,
in Melbourne, considerable observational and theoretical progress has
been made in our understanding of the first two of these, while
considerable technical progress has been made in the third.
\end{abstract}

\section{Confession}

Predicting whether a conference will prove exciting or ho-hum has
always been difficult for me.  So it is with especially deep
admiration that I congratulate the Scientific Organizing Committee on
the superb job they have done (at least until this point) in selecting
speakers for this meeting.  We have been treated to an excellent
smorgasbord of reviews, background talks and exciting new results.
But on second thought, perhaps the SOC did not have so demanding a job
after all.  The poster presentations have been so outstanding, with
such very high signal to noise, that a more or less random selection
from among the submitted abstracts might have produced an equally good
set of talks.

\section{Question}

This said, let me pose a question which may seem churlish: why did we
have this meeting?  After all, the cost of such a gathering is very
considerable, especially in person hours spent preparing for it and in
actual attendance, but also in the cost of travel and discomfort --
including circling over Logan airport for several hours, being
diverted to Hartford, New York or Washington, and being drenched in
downpours.

We've heard ourselves referred to by several of our speakers as ``the
lensing community."  If there really is such an entity, it is one that
spontaneously fragments into constituencies. There is a microlensing
community and there are strong and weak lensing communities, with the
latter again split into those who study weak lensing by identifiable
objects and those who study the properties of the potential field
without little regard for the objects responsible for it.

Were you to look at a list of astronomical meetings in any given year,
you would find that they fall into two broad categories.  The majority
of meetings are organized around some class of astronomical phenomenon:
peculiar A stars, high redshift galaxies, molecular clouds.  But many
meetings are organized around specific techniques rather than
phenomena: long baseline interferometry, adaptive optics, TeV
astronomy.  Typically these are in areas where the technology is new
or rapidly changing.  And so there is a second question related to
my first question.  Is this meeting a phenomenological meeting or is it a
technological meeting?

The list of the phenomena covered by participants at this meeting is
very long.  One is impressed at how large a fraction of the
astronomical universe has been discussed: planets, stellar surfaces,
quasars and their hosts, the microwave background.  Nobody would ever
organize a meeting with so wide a range of topics; should we conclude,
by elimination, that this is a technique meeting?

\begin{figure}
\vspace {4.0 truein}
\includegraphics{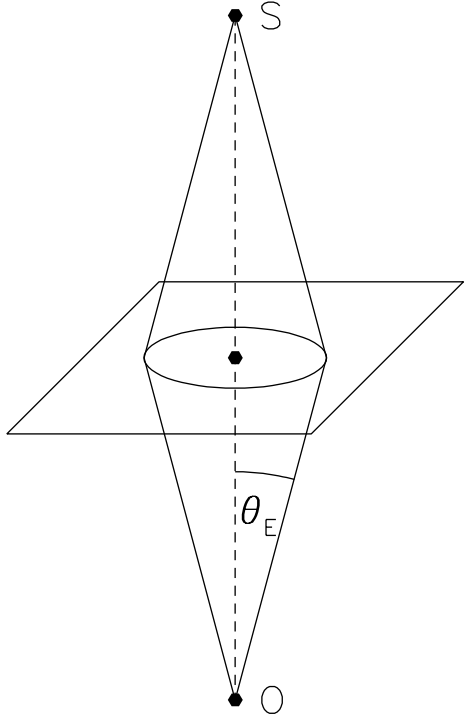}
\caption{The ``classic'' lens diagram, showing source, observer, and lens plane (courtesy
of M. Bartelmann).}
\end{figure}

In addressing this question it is helpful to look at the classic
gravitational lensing diagram, the variants of which Virginia {\sc
Trimble} traced back over two hundred years.  The figure has three
sections -- the source end, the observer end, and middle, where the
lensing takes place.  Phenomenon oriented meetings are usually
concerned with the source end of the diagram.  Technique oriented
meetings are usually concerned with the observer end of the diagram.
What makes this meeting unusual, distinguishing it from most other
meetings, is that the principal interest of most contributors at this
meeting has been neither at the observer end of the diagram nor at the
source end but the machinery in the middle.  We would seem to be in a
class by ourselves (though not quite if we count students of the
interstellar medium).

The subject of our meeting goes by three different names, each of
which carries somewhat different connotations.  In French it is
called ``mirage gravitationnel," which tends to emphasize the
experience of the observer.  Some of our speakers describe their use
of ``gravitational telescopes."  For them the effect is simply a tool
which gives them more photons or higher resolution from an otherwise
faint or small source.  The term ``gravitational lens" emphasizes the
intermediate optics rather than the astronomical sources or the
observer.

If I were to characterize where our contributors' interests lie, I
would guess that 25\% are primarily interested in the sources and
15\% are primarily interested in the detectors and analysis
techniques with the remaining 60\% interested in the lenses
themselves, though of course there's scarcely a person in this room
who isn't interested in all three.

The reason for interest in the lenses is manifest -- observations
of the deflections, distortions and delays introduced by lensing
permit one to measure the masses of intervening objects.  The
circumstances under which astronomers can measure masses are so rare
that they jump at the opportunity.

The history of the measurement of masses of clusters of galaxies
drives this point home.  Zwicky and Smith were the first to measure
the mass of a cluster, but the answer they found was so far from the
expectations of the day that the astronomical community chose to
ignore it.  By the mid-1970s measurements of X-ray gas profiles and
temperatures more or less confirmed the optical velocity dispersion
measurements, but the astronomical community was still in a state of
denial about the consequences.  Doubts and suspicions have lingered
into the present, so people have seized upon gravitational lensing as
a means of resolving the issue.

Our classic lensing diagram, as drawn here, is grossly exaggerated,
and represents a rather bad case of wishful thinking (something to
which we, as astronomers, are in no way immune).  Our figure pretends
to be a case of strong imaging.  The widest separations, of order 1
arcminute -- a third of a milliradian -- are produced by clusters of
galaxies.  Strong lensing by galaxies produces deflections of several
microradians.  Microlensing within the Local Group produces
deflections of nanoradians, and microlensing on cosmological scales
gives deflections measured in picoradians.  Even for the largest
deflections we consider, this diagram collapses to a line if one tries
to draw it to scale.  The exceedingly small solid angle of influence
of the lenses we seek to study drives us to extremes in terms of
photometric accuracy, astrometric precision and in numbers of objects
needed to produce the wanted effects.  In some cases that quest
borders on the quixotic.  The fact that so many of us are willing to
undertake such efforts is testimony to how important we believe the
results might be.

I will use our classic diagram in reviewing what we've heard and
seen in the last five days, grouping together results at the observer
end, results at the source end and results in the middle.

\section{The Observer}

The lensing community can take considerable pride in the extent to
which some of our members have led the larger astronomical community
in experimental techniques.  Chief among these has been the
development large format CCD detectors.  When Tony Tyson first
undertook measurements of galaxy-galaxy lensing in the early 1980s, he
used photographic plates.  We have all had a good laugh at the
old-fashioned darkroom timer that was called out of retirement to keep
our speakers in line.  But remember that well into the 1990s, the
photographic plate, despite its 1\% quantum efficiency and its
horribly non-linear response, has remained a valuable tool in our
field because of the small size of solid state detectors.  The MACHO
and EROS groups, Tyson and Bernstein, and more recently Gerry Luppino
and his group have been world leaders in constructing large area solid
state imagers.

A number of gravitational lens programs has been very large in scale,
requiring a degree of organization and coordination rarely seen in
astronomy.  The MACHO and EROS collaborations, in particular, have
brought the culture of particle physics to ground based astronomy.
The CLASS collaboration ({\sc Myers)}, the MIT surveys ({\sc Winn})
and the ACT effort ({\sc Prouton}) have brought a new style to radio
observations as well, with radio telescopes spending almost as much
time slewing between objects as observing.

I wish I could say that optical astronomers have done as good a job as
radio astronomers in searching for new lenses.  Strong lensers (myself
among them) have not been as effective in marshalling the resources
necessary.  There is a crying need for wide field optical telescopes
of moderate size with silicon focal planes to carry out survey work.
On a more positive note, the PLANET and G-MAN collaborations have been
spectacularly successful in assembling the instruments necessary to
carry out round the clock monitoring of exotic lensing events.

Lensers have also lead the way in software.  FOCAS was an early effort
on the part of Tyson and his collaborators (Jarvis {\it et al.} 1981)
to deal with unprecedentedly large numbers of images.  The nearly
total automation of the MACHO project may not seem particularly
noteworthy to radio or X-ray astronomers, but it is quite remarkable
among ground-based optical efforts.  The OGLE program, we are told, is
automated to the point where a program field is specified and the
reduced data are emailed to a designated recipient.  Image
differencing is another development which, while straightforward in
principle, has only now been made to work, and which promises major
improvements in sensitivity.  It is remarkable that {\sc Crotts} and
his collaborators and now Alard and Lupton (1998) have been able to press to
the photon limit.

Alas one must worry not only about photon statistics but also
systematic errors.  The efforts of the various weak lensing groups to
remove myriad sources of systematic image distortion have been nothing
short of heroic.  Chris {\sc Fassnacht} and Leon {\sc Koopmans} have
likewise pushed the envelope in their exceedingly accurate radiometric
measurements.  We have also seen extraordinarily high dynamic range
measurements in the ring of B0218+357 ({\sc Biggs}) which will help in
its modelling.  The UH optical astronomers are the first I know of who
have dared to show the Hubble Deep Field image side by side with their own.
Theirs may be somewhat less deep, but it is certainly very much
wider, and that is clearly what we need for weak lensing.

Some measure of the excitement generated by the phenomenon of
gravitational lensing can be had by noting the prominent role it plays
in the justification for many of the major projects now being
evaluated by the US National Research Council's decennial survey of
astronomy.  We have heard lensing invoked as a justification for NGST,
for an 8m ground based ``dark matter telescope", for the VLA+ upgrade
and for the Square Kilometer Array.  Lensing likewise figures
prominently in the programs for the Advanced Camera for Surveys on
HST, Chandra, XMM, SIM and Planck.

\section{Sources}

Lensing has provided data about sources which could not otherwise have been
obtained.  Hans-Walter {\sc Rix} has shown us that the hosts of high
redshift quasars are surprisingly easy to see if one uses lensing
to boost the resolution of NICMOS.  It is not atypical to find
that the increased resolution produced by a lens is more important than
the increased photon count.

A number of speakers and presenters have shown us how microlensing can
be used to set limits on the sizes of quasar components ({\sc Agol;
Yonehara)} both in the optical and in the radio.  Some of our
theorists have outlined how one might use a caustic moving across a
quasar to study the structure of quasar accretion disks.

We have heard from Penny {\sc Sackett} and others about how
microlensing can be used to study the surfaces of stars, and in
particular to check models of limb darkening and for the presence of
starspots.  Such stars have diameters (if I heard correctly) of 100
nanoseconds.

Bob {\sc Nemiroff} told us how gravitational lenses give us information about
otherwise elusive gamma ray bursts.  It should be noted that one of the two
subclasses of gamma ray bursts, the short ones, have not yet had host
galaxies measured.  Lensing may therefore provide the only limit on the
redshifts of these objects, albeit a weak one.

And we have heard, {\it en passant}, about how gravitational lensing has
twice given us the record holding high redshift galaxies, first in a
CNOC cluster (Yee {\it et al.} 1996) and then in 1358+62 (Franx {\it et al.}
1997).

It is notable that we have {\it not} had at talk about the use
of gravitational telescopes to improve the spatial resolution of the
submillimeter bolometer array (SCUBA) on the JCMT in the study of dusty
galaxies at high redshift.  Some of the best work in that field has
been done using lensing (e.g. Smail {\it et al.} 1997), and the people
who did it have in years past been active participants at
gravitational lens meetings.  I doubt that our SOC slighted this work;
rather, I suspect that these individuals treat gravitational
telescopes as just another weapon in the astronomical armory, and that
they feel their time is more wisely spent going to meetings where the high
redshift universe is the principal focus.

\section{The Lenses}

The majority of the papers at this meeting have emphasized neither the
sources nor the observing and detection but the deflection, distortion and
delay of light by intervening masses.  Until now the masses of astronomical
objects have been measured by observing the bound orbits of gas, stars or
galaxies in gravitational potentials.  Now we study mass distributions by
studying the unbound the orbits of photons.  Until now we have relied there
being stars, gas or galaxies present to study potentials.  But now we know,
at least on average, that we can count on a certain number of background
sources to be lensed by the foreground objects we wish to study.

Lensing might not be quite so interesting were the universe not pervaded by
dark matter.  We suspect that 90\% or more of the matter in the universe is
non-baryonic, and that a major fraction of the baryonic matter may itself be
dark (though not quite so totally and unrelentingly dark as the non-baryonic
stuff).  It is with a combination of embarrassment and frustration that we explain
to people outside astronomy that we cannot observe 90\% of the universe.
Gravitational lensing offers us a chance to redeem ourselves.

What our friends outside astronomy don't know is that luminous matter
is at best a treacherous tracer of dark matter.  We know that light
fails to trace mass in the Milky Way and other galaxies, and we
suspect that galaxies may fail to trace light in bound clusters of
galaxies and in yet larger structures in the universe.  So we are
driven to gravitational lensing as the most reliable means of studying
the distribution of dark matter.

There has been spirited discussion of the observation of gravitational
microlensing toward the Magellanic clouds and its implication for the
composition of the dark halo of the Milky Way.  While the MACHO
collaboration has argued that most of these events arise from compact
objects in the galactic halo (Alcock {\it et al.} 1997), we have heard
forceful arguments for self lensing by the LMC ({\sc Evans}).  Given
theorists' creativity in coming up with models, the issue is likely to
be settled only with a very much larger set of events than we
presently have.  However the question is resolved, we will have
learned an enormous amount from the microlensing searches.

Another subject which generated fascinating discussion was the
gravitational potentials of galaxies for which time delays have been
measured.  There has been superb progress on the observational front.
At the Melbourne IAU symposium (Kochanek and Hewitt 1996) even the
time delay for B0957+561 was a matter of contention.  Today there are
8 systems with measured time delays, with two of those delays reported
for the first time at this meeting.  This is the result of prodigious,
painstaking effort on the part of radio and optical observers.  It is
easy to forget that a set of 50 data points demands 50 times the
effort (perhaps even more, given the spacing requirements) than a
single data point.  The first reported delays for RX J0911+0551 and
CLASS B1600+434, from data obtained by Ingunn {\sc Burud} and
collaborators with the NOT, were breathtaking.  Tommy {\sc Wicklind's}
confirmation of the time delay for PKS B1830-211 using single dish
molecular absorption spectra was another {\it tour de force}.

There are several schools regarding the interpretation of time delays.
There are those who choose parameterized models for potentials ({\sc
Bernstein; Chae}) and those who despair of adequate parameterization
and instead adopt a non-parametric approach ({\sc Saha; Williams}).  There are
those who insist that every detail of the gravitational potential
(most importantly its logarithmic slope) must be measured from the
lens itself.  On the other hand are those who are willing to bring
their knowledge of the dynamics of other galaxies to bear on the
problem.  The former are the perfectionists, members of the ``golden
lens" school.  The latter are the compromisers, members of the ``warts
and all'' school.  As a member of the latter, I will exercise my
prerogative as summarizer and note that if one adopts a simple model
and observes a small scatter in the derived values of the Hubble
constant, one might not be making so large an error in transferring
one's hard won knowledge of galaxy dynamics to galaxies for which the
dynamics are nearly impossible to measure.  In this regard my reaction
to Liliya {\sc Williams'} non-parametric models was exactly the
opposite of Roger {\sc Blandford's}.  Where he drew the conclusion
that the Hubble constant was hopelessly uncertain, I was pleased to
see how little the Hubble constant depended on anything except the
logarithmic slope of the potential, a result also emphasized by Olaf
{\sc Wucknitz}.

In his review, Ed {\sc Turner} opined that lenses now give the best
value of the Hubble constant.  Considering the care that has gone into
the HST Cepheid Key Project, especially in estimating their error budget, I
don't think we are yet in a position to claim this particular piece of
high ground.  But if we see redshifts for RX J0911+0551 and HE
B1104--1805, and if in another year the present time delay results
don't change dramatically, it might be that even unbiassed observers
(creatures rarer than unicorns) would agree with him.

Both those of the ``golden lens'' school and those of the ``warts and all''
school agree that many new lenses are needed.  Survey work is {\it
sine qua non} of astronomy.  CLASS ({\sc Browne; Myers; Rusin}) has been
gloriously successful in producing new cases, including two of those
for which we now have delays.  Optical searches have until now lagged
behind, especially when one folds in the fact that 90\% of quasars are
radio quiet.  The Sloan telescope in the north ({\sc Pindor}) and the
VST in the south may go part way toward redressing this imbalance, but for
reasons which are in no way fundamental (e.g. pixel size, programatic
constraints) neither is ideally suited to the task of finding strong
lenses.

The strong lenses have also given us a picture of the luminosity
evolution of early type systems which is completely independent of the
work done in clusters of galaxies.  It is amazing that the results
reported by {\sc Kochanek}, determining parameters for the so-called
``fundamental'' plane using lensing galaxies, agree as well as they do
with results obtained for clusters using conventional methods.  Who
would have thought that galaxies selected by mass would agree as well
as they do with galaxies selected by light?

Brian {\sc McLeod} spoke about the non-gravitational aspects of
propagation of multiply imaged quasar light through lens galaxies,
giving us a unique handle on the properties of the ISM at high
redshift.  In the course of that he was able to show that, for
whatever mysterious reason, lensing has helped us to find two of the
intrinsically reddest quasars known in the universe.

It must be remembered that strong gravitational lenses are poorly
designed and, moreover, fabricated from inferior materials.  The lens
material typically exhibits huge variations in its index of refraction
due to the substantial percentage of its mass in stars and MACHOs.
The stars {\it must} introduce microlensing even if MACHOs do not.  Here
again we've begun to address questions which I would not have thought
possible.  While there is a near degeneracy between the rms mass of
the microlenses and the fraction of the intervening mass in condensed
objects, there is hope for separating these two effects in the higher
order statistics of light curves.  One need only remember that Sjur
{\sc Refsdals's} two curves, one based on a peak and the other based
on a plateau, did not have the same shape in his log-log ``exclusion''
diagram. {\sc Koopmans'} results on CLASS B1600+434 are all the more
fascinating for being a case of observation not yet confirmed by
theory.  While microlensing is noise to those who wish to measure time
delays, perhaps we must count ourselves lucky that at least some of
our lenses suffer from it.

The developments in galaxy-galaxy lensing have been very encouraging.
Several groups have described their efforts ({\sc Brainerd; Casertano;
Fischer; Smith}) and, quite remarkably, they all agree with each
other.  We still haven't seen the cutoff expected in our isothermal
sphere models and Hank {\sc Hoekstra's} result for groups lead me to
suspect that we may never see one.  But there are other things to be
tried, including testing of the isothermality hypothesis.  Phil {\sc
Fischer} showed that variations in the Sloan survey PSF were not so
malignant as to swamp the galaxy-galaxy lensing signal.

Probably only at meetings on adaptive optics do point spread functions
receive more attention than they did at this one.  Hans-Walter {\sc
Rix} described the NICMOS PSF as one that only a mother could love.  I
suspect these words will find new application as people analyze the
data obtained with new generations of wide field cameras now coming on
line.

Weak lensing observations of clusters have moved from the regime of
mar-ginal detection to that of serious astrophysical tool.  Nick {\sc
Kaiser} has shown us that there is surprisingly little radial bias in
the luminosity profiles of clusters of galaxies -- this from the man
whose name is most closely associated with the concept.  It's far too
soon to accede on this point -- there are troubling differences
between lensing results and those obtained from optical and X-ray
data.  I wonder whether we shouldn't introduce a few weak lensing
``standards'', in the same way we have adopted photometric standards,
to ensure that everyone is on the same system.  A point that was made
many times, which may nonetheless have failed to penetrate the
stubbornest of listeners, is that ``seeing is everything."

With the successful launch of Chandra and the promise of XMM and
several CMB imagers, we have the prospect of comparing 4 different
estimators of mass and substructure in clusters of galaxies.  A
crucial issue in this regard is the boundary between galaxy and
cluster -- where the galaxy ends and the cluster begins.  Priya {\sc
Natarajan's} results have whetted our appetite for further
investigations of this sort.

On the scales so large that structure is still linear or weakly
non-linear, scales on which one must study fields rather than objects,
we have heard about mean polarizations ({\sc Wilson}) and polarization
correlations ({\sc Wittmann}) and aperture masses ({\sc Schneider}) as
alternative vehicles for studying the the power spectrum of mass
fluctuations.  The complementarity (a word that brings down the duck
with \$100) of weak lensing results and CMB measurements has been
repeatedly emphasized as has the point that these probe large scale
structure at different epochs ({\sc Jain; Seljak}).  It is a measure
of how exciting these prospects are that people are willing to
undertake huge programs of extraordinarily delicate measurement.  The
signal may already have been seen but the uncertainties, almost all of
them systematic, are as yet too poorly understood for firm conclusions
to be drawn.

Next there is the small matter of the universe itself.  In addition to
the dimensioned parameter $H_0$, lensing can in principle tell us
about dimensionless parameters, the cosmological density parameter
$\Omega_m$, and the dimensionless version of the cosmological constant
(or the vacuum energy density), $\Omega_\Lambda$.  There are several
approaches to measuring these.  The largest effect involves the
numbers of lensed systems ({\sc Helbig}), but as yet the luminosity
functions for unlensed objects and the mass functions (and shapes) of
lenses are too poorly known for these to provide strong limits.  There
are other effects, such as comparison of the sizes of Einstein rings
for objects at different redshifts behind a lens ({\sc Link}).  We may
get lucky in this regard and find a lens with simple geometry and
multiple sources each multiply imaged.

Finally let's return to our own neighborhood and consider a different
kind of dark matter -- planets.  We have seen that planet detection is
a serious possibility ({\sc Dalil; Gaudi}) and will be even more
likely with the advent of SIM ({\sc Boden}).  We enjoyed an outlaw
poster claiming a planet of 5 $M_J$ has already been observed in a
microlensing event (Bennett {\it et al.} 1999).

\section{Broad Issues}

Several themes emerged in the course of the meeting which don't fit
easily into our observer, source or lens pigeonholes.  The first of
these regards a sea change in the way we do astronomy.  Many of you
saw the article in Sunday's {\it NY Times Magazine} called ``The
Loneliness of the Long Distance Cosmologist" (Panek, 1999).  It
describes how the nature of the astronomical enterprise in general,
and how measurement of $H_0$ in particular, has changed from the solo
effort of a lone wolf carried out at the prime focus of a unique
telescope to the concerted effort of a large team, often using
multiple telescopes (which many members of the team may never even
have have seen).  While there may still be room for lone wolves in
gravitational lensing, team efforts, with the attendant headaches, the
massaging of egos and the compromising on means and ends, seems to be
the order of the day.  Like it or not, we are destined to live in an
era of cloying and lame acronyms.

A second recurring theme, not unrelated to the first, has been that of
the ``exclusion'' diagram.  We have seen many instances of
observations that, while they rule out the large volumes of model
space, produce allowed volumes (error ellipsoids, to first order)
whose principal axes happen not to lie parallel to the axes of the
model space of interest.  The lone wolves among us show a strong
preference for measurements which produce nicely aligned error
ellipsoids.  The team players don't care whether ellipsoids (singly or
from more than one measurement) and axes are aligned or not, as long
as the resulting volume is small.  I can sympathize with the lone
wolves on aesthetic grounds but the future belongs to the tilted
ellipsoids.

\section{Bread and Butter Issues}

The success of the Scientific Organizing Committee has been surpassed
only by that of the Local Organizing Committee.  With the exception of
a friendly visit by the local firefighters our meeting has proceeded
seamlessly.  The accommodations have been excellent, the meeting room
and poster area ideal, the pastry and coffee far above average, and
the dinner cruise up and down the Charles and around a moonlit Boston
harbor most memorable.  We owe the chair of the LOC, Tereasa Brainerd,
our host institution, Boston University, and our sponsors, the US
National Science Foundation, NASA, and Boston University, considerable
thanks for making this meeting as productive as it has been.

Perhaps the most appropriate place to end is with a call for
volunteers to organize a gravitational lens meeting in 2002!


\begin{references}
\reference Alard, C. and Lupton, R.H. 1998 \apj, 503, 325 
\reference Alcock, C., {\it et al.} 1997 \apj, 486, 697
\reference Bennett, D.P., Rhie, S.H. {\it et al.} 1999,
{\it astro-ph/9908038}
\reference Franx, M., Illingworth, G.D., Kelson, D.D.,
van Dokkum, P.G. and Tran, K. 1997 \apjl, 486, L75
\reference Jarvis, J.F., Tyson, J.A. 1981 \aj, 86, 476
\reference Kochanek, C.S. and Hewitt, J.N. 1996 {\it IAU Symposium 173:
Astrophysical Applications of Gravitational Lensing} 
\reference Panek, R. 1999 {\it NY Times Magazine}, 25 July, p.22
\reference Smail, I., Ivision, R.J., and Blain, A.W. 1997 \apjl, 490, L5
\reference Tyson, J.A., Valdes, F., Jarvis, J.F., Mills, A.P., Jr.
1984 \apjl, 281, L59  
\reference Yee, H., Ellingson, E., Bechtold, J., Carlberg, R. and
Cuillandre, J. 1996 \aj, 111, 1783
\end{references}
\end{document}